\documentclass[aps,prd,12pt,notitlepage,showpacs,nofootinbib,tightenlines]{revtex4-1}
\usepackage{amsmath}
\usepackage{amssymb}
\usepackage{epsfig}
\usepackage{graphicx}
\usepackage{bm}
\usepackage{times}
\usepackage{braket}
\usepackage{color}
\usepackage{slashed}
\usepackage{hyperref}

\definecolor{Red}{rgb}{1.,0.,0.}

\definecolor{Blue}{rgb}{0.,0.,1.}

\definecolor{nicered}{rgb}{0.7,0.1,0.1}
\definecolor{nicegreen}{rgb}{0.1,0.5,0.1}
\hypersetup{colorlinks,citecolor=nicegreen,linkcolor=nicered}

\begin{document}
%%%%%%%%%%%%%%%%%%%%%%%%%%%%%%%%%%%%%%%%%%%%%

\newcommand{\beq}{\begin{eqnarray}}
\newcommand{\eeq}{\end{eqnarray}}

\newcommand{\non}{\nonumber\\ }
\newcommand{\ov}{\overline}

\newcommand{\calo}{ {\cal O}}
\newcommand{\calb}{ {\cal B}}
\newcommand{\calr}{ {\cal R}}

\newcommand{\rmt}{ {\rm T}}
\newcommand{\jpsi}{J/\psi}
\newcommand{\etab}{\bar{\eta} }
\newcommand{\etar}{\eta^\prime }
\newcommand{\etap}{\eta^{(\prime)}}
\newcommand{\psl}{ p \hspace{-2.0truemm}/ }
\newcommand{\qsl}{ q \hspace{-2.0truemm}/ }
\newcommand{\epsl}{ \epsilon \hspace{-2.0truemm}/ }
\newcommand{\nsl}{ n \hspace{-2.2truemm}/ }
\newcommand{\vsl}{ v \hspace{-2.2truemm}/ }

%%%%%%%%%%%%%%%%%%%
\def \cpc{{ Chin. Phys. C } }
\def \ctp{{ Commun. Theor. Phys. } }
\def \epjc{ Eur. Phys. J. C }
\def \jhep{ JHEP  }
\def \jpg{ J. Phys. G }
\def \mpla{ Mod. Phys. Lett. A  }
\def \npb{  Nucl. Phys. B }
\def \plb{ Phys. Lett. B }
\def \ppnp{ Prog.Part. $\&$ Nucl.Phys. }
\def \pr{  Phys. Rept.  }
\def \prd{  Phys. Rev. D }
\def \prl{  Phys. Rev. Lett.  }
\def \ptep{ Prog.Theor. Exp. Phys.   }
\def \ptp{  Prog. Theor. Phys. }
\def \zpc{  Z. Phys. C}
\def \csb{  Chin. Sci. Bull.  }
%%%%%%%%%%%%%%%%%%%%%%%%%%%%%%%%%%%%%%%%%%%%%%%%%%%%%%%%%%%
\title{The $S$-wave resonance contributions to the three-body decays $B^0_{(s)}\to
\eta_c f_0(X)\to \eta_c\pi^+\pi^-$ in perturbative QCD approach}
\author{Ya Li$^1$}\email{liyakelly@hotmail.com}
\author{Ai-Jun Ma$^1$}  \email{theoma@163.com}
\author{Wen-Fei Wang$^2$} \email{wfwang@sxu.edu.cn}
\author{Zhen-Jun Xiao$^{1}$}\email{xiaozhenjun@njnu.edu.cn}
\affiliation{1. Department of Physics and Institute of Theoretical Physics,
                Nanjing Normal University, Nanjing, Jiangsu 210023, P.R. China}
\affiliation{2. Institute of Theoretical Physics and Department of Physics,
                Shanxi University, Taiyuan, Shanxi 030006, China}
\date{\today}
%-----------------------------------------------------%
\begin{abstract}
In this paper, we study the three-body decays $B^0/B^0_s \to \eta_c f_0(X)\to \eta_c \pi^+\pi^-$ by employing
the perturbative QCD (PQCD) factorization approach. We evaluate the $S$-wave resonance contributions by using
the two-pion  distribution amplitude $\Phi_{\pi\pi}^{\rm S}$. The Breit-Wigner formula for the $f_0(500)$,
$f_0(1500)$, and $f_0(1790)$ resonances and the Flatt\'e model for the $f_0(980)$ resonance are adopted to
parameterize the time-like scalar form factors $F_{s}(\omega^2)$.
We also use the D.~V.~Bugg model to parameterize the $f_0(500)$ and compare the relevant theoretical predictions from
different models. We found the following results:
(a)  the PQCD predictions for the branching ratios are  ${\cal B}(B^0\to \eta_c f_0(500)[\pi^+\pi^-])=
\left ( 1.53 ^{+0.76}_{-0.35} \right ) \times 10^{-6}$ for Breit-Wigner model and ${\cal B}(B^0\to \eta_c f_0(500)[\pi^+\pi^-])=
\left ( 2.31 ^{+0.96}_{-0.48} \right ) \times 10^{-6}$ for D.~V.~Bugg model;
(b) $ {\cal B}(B_s\to \eta_c f_0(X)[\pi^+\pi^-] ) =\left ( 5.02^{+1.49}_{-1.08} \right )\times 10^{-5}$
when the contributions from $f_0(X)=(f_0(980),f_0(1500),f_0(1790))$ are all taken into account;
and (c) The considered decays could be measured at the ongoing LHCb experiment, consequently,
the formalism of two-hadron distribution amplitudes could also be tested by such experiments.
\end{abstract}


\maketitle
%------------------------------------------------------%

\section{Introduction}\label{sec:1}

Several years ago, some three-body hadronic $B \to 3h^{(')}$ ( $h,h'=\pi, K$) decays have been measured by
BaBar and  Belle Collaborations \cite{bf1} and studied by using the Dalitz-plot-analysis.
The LHCb Collaboration reported, very  recently,  their experimental measurements for
the branching ratios and sizable direct CP asymmetries for  some three-body charmless hadronic decays
$B^+ \to K^+ K^+K^-, K^+ K^+\pi^-, K^+ \pi^+\pi^-$ and $\pi^+ \pi^+\pi^-$ \cite{Aaij:2013sfa,Aaij:2013bla,1608a},
the three-body charmed hadronic decays $B^0\to \jpsi \pi^+\pi^-$ \cite{lhcb01,lhcb02,lhcb03,Aaij:2014emv} and
$B^+\to \jpsi \phi K^+$ \cite{1606a},
or the decays $B_s^0\to \jpsi K^+K^-$, $\jpsi \phi\phi$, $\bar{D}^0 K^- \pi^+$ \cite{lhcb02,lhcb03,jhep16a}.
The large localized CP asymmetry observed by
LHCb brings new challenges to experimentalists and their traditional models to fit data, and also has invoked
more theoretical studies on how to understand these very interesting three-body $B/B_s$ meson decays.

On the theory side, the three-body hadronic decays of the heavy $B/B_s$  meson
are much more complicated to be described theoretically than those two body $B/B_s \to h_1 h_2$
(here $h_i$ refer to light mesons $\pi, K, \rho, etc$  ) decays.
During the past two decades, such two-body hadronic $B/B_s$ meson decays  have been studied
systematically and successfully by employing various kinds of factorizations approaches.
The three major factorization approaches are
the QCD-improved factorization (QCDF) \cite{bbns99,npb675,beneke2007}, the perturbative QCD (PQCD) factorization
approach \cite{Keum:2000ph,li2003,xiao-1a,xiao-1b,ali2007-bs} and the soft-collinear-effective theory (SCET)
\cite{scet01,scet02}.
For most $B/B_s \to h_1 h_2$ decay channels, the theoretical predictions obtained by using these different
factorization  approaches agree well with each other and also be well consistent with the data within
errors.

For $B/B_s$ three-body hadronic decays, however, they do receive both the resonant and non-resonant contributions,
as well as the possible final state interactions (FSIs), while the relative strength of these contributions
are varying significantly for different decay modes.
They are known experimentally to be dominated by the low energy resonances on $\pi\pi$, $KK$ and $K\pi$ channels
on Dalitz plot, usually analysed by employing the isobar model  in which the decay amplitudes are parameterized by
sums of the Breit-Wigner terms and a background, but without the inclusion of the possible
contributions from the coupled channels and the three-body effects such as the FSIs.
In fact, the three-body hadronic $B/B_s$ meson decays have been studied for many years for example  in
Refs.~\cite{Chen:2002th,Chen:2004az,CY02,FPP04,65-094004,79-094005,cheng13,cheng14,cheng16,liy2015,BIL14,NR11,MN11,npb899-247}
by employing the isobar model and/or other rather different theoretical approaches, but it is still in the early stage
for both the theoretical studies and the experimental measurements of such kinds of
three-body decays. For example, the factorization for such three-body decays has not been verified yet,
and many important issues remain to be resolved.

In Ref.~\cite{79-094005}, for instance, the authors studied the decays of $B \to K \pi \pi$ by assuming the validity
of factorization for the quasi-two-body  $B \to (K \pi)_{S,P} \pi \to K\pi \pi$ and introducing the scalar
$f_0^{K\pi}(q^2)$ and vector $f_1^{K,\pi}(q^2)$ form factors to describe the matrix element
$<K^- \pi^+|(\bar{s} d)_{V-A}|0>$.
From the viewpoint of the authors of Ref.~\cite{65-094004}, a suitable
scalar form factor could be developed by the chiral dynamics of low-energy hadron-hadron
interactions, which is rather different from the Breit-Wigner form adopted to study $B \to \sigma \pi$.

In Refs.~\cite{cheng13,cheng14,cheng16}, %% R19=PRD88-114014
the authors calculated the branching ratios and direct CP violation for the
charmless three-body hadronic decays $B \to 3h$ with $h=(\pi,K)$ by using a simple model based on the factorization approach.
They evaluated the non-resonant contributions to the considered decays in the
framework of heavy meson chiral perturbation theory (HMChPT) with
some modifications, while describing the resonant contributions by using the isobar model in terms of  the usual Breit-Wigner
formalism. The strong phase $\phi$, the parameter $\alpha_{\rm NR}$ and the exponential
factor $e^{-\alpha_{\rm NR} p_B\cdot (p_i+p_j)}$ are introduced
in their works \cite{cheng16} in order to accommodate the data.

In PQCD factorization approach, however, we study the three-body hadronic decays of B meson by introducing
the crucial non-perturbative input of the two-hadron distribution amplitude (DA) $\Phi_{h_1 h_2}$ \cite{MP} and use the
time-like form factors to parameterize these two-hadron DAs.
In our opinion, a direct evaluation of hard $b$-quark decay kernels, which contain two virtual gluons at
leading order (LO), is power-suppressed and not important.
When there is at least one pair of light mesons having an invariant mass below $O(\bar\Lambda m_B)$
\cite{Chen:2002th} ( here $\bar\Lambda=m_B-m_b$ being the $B$ meson and $b$ quark mass difference),
the contribution from this region is dominant.
The configuration involves two energetic mesons almost collimating to each other, in which three-body
interactions are expected to be suppressed.
However, the relative importance of the contributions from the two hard gluon exchanges and from the
configuration with two collimating mesons still depend on specific decay channels and
kinematic regions considered.
It seems reasonable that the dynamics associated with the
pair of mesons can be factorized into a two-meson distribution amplitude $\Phi_{h_1h_2}$ \cite{MP}.
One can describe the typical PQCD factorization formula for a $B\to h_1h_2h_3$ decay
amplitude as the form of ~\cite{Chen:2002th,Chen:2004az}
\beq
\mathcal{A}=\Phi_B\otimes H\otimes \Phi_{h_1h_2}\otimes\Phi_{h_3},
\eeq
where the hard kernel $H$ describes the dynamics of the strong and electroweak interactions in three-body
hadronic decays in a similar way as the one for the two-body hadronic $B\to h_1 h_2$ decays,
the functions $\Phi_B$,  $\Phi_{h_1h_2}$ and $\Phi_{h_3}$ are the wave functions for the B meson
and the final state mesons, which absorbs the non-perturbative dynamics in the process.
Specifically, $\Phi_{h_1h_2}$ is the two-hadron ($h_1$ and $h_2$) DAs proposed for example in
Refs.~\cite{MP,MT,MN}, which describes the structure of the final state $h_1-h_2$ pair, as illustrated
explicitly in Fig.~\ref{fig:fig1}.

By employing the PQCD approach, the authors of Ref.~\cite{Wang-2014a} studied the $B^\pm \to
\pi^\pm \pi^+ \pi^-$ and $ K^\pm \pi^+\pi^-$ decays, evaluated the direct $CP$ asymmetries
by fitting the time-like form factors and the rescattering phases contained in the two-pion
distribution amplitudes to relevant experimental data, the resulted PQCD predictions
agree well with the LHCb measurements~\cite{Aaij:2013sfa,Aaij:2013bla}.
In Ref.~\cite{Wang-2014a}, however,  only the non-resonant contributions to the time-like form factors were taken
into account, the regions involving intermediate resonances are not considered.
In the new work ~\cite{Wang-2015a}, by parameterizing the complex time-like form factors
which include both resonant and non-resonant contributions, the authors studied the three-body decays
$B_s\to J/\psi f_0(980)[f_0(980)\to\pi^+\pi^-]$ and
$B_s\to f_0(980)[f_0(980)\to\pi^+\pi^-]\mu^+\mu^-$ decays by using the $S$-wave two-pion DAs.

In recent years, significant improvements for understanding the heavy quarkonium production mechanism
have been achieved ~\cite{Aaij:2015cs}. The meson $\eta_c$ and $J/\psi$ have same quark content but with
different spin angular momentum.
Following Ref.~\cite{Wang-2015a}, we here will study the three-body hadronic decays $B^0_{(s)}
\to \eta_c f_0\to \eta_c [f_0\to \pi^+\pi^-]$.
We will consider the $S$-wave resonant contributions to the decay $B^0\to \eta_c f_0(500)\to \eta_c
(\pi^+\pi^-)$, as well as the decays $B^0_s\to \eta_c f_0(X)\to \eta_c (\pi^+\pi^-)$
with $f_0(X)=(f_0(980),f_0(1500),f_0(1790))$.
Apart from the leading-order factorizable contributions, we also take into account the NLO vertex corrections
to the Wilson coefficients.
In Sec.~II, we give a brief introduction for the theoretical framework and present the expressions
of the decay amplitudes. The numerical values, some discussions and the conclusions will be given in last two sections.

\section{The theoretical framework}\label{sec:2}

\begin{figure}[tbp]
\vspace{-1cm}
\centerline{\epsfxsize=14cm \epsffile{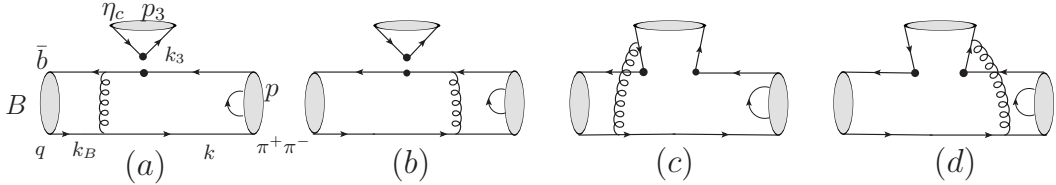}}
\caption{Typical Feynman diagrams for the three-body decays $B^0_q \to \eta_c\pi^+\pi^-$ with $q=(d,s)$,
and the symbol $\bullet$ denotes the weak vertex.}
\label{fig:fig1}
\end{figure}

By introducing the two-pion DAs, the $B^0_{(s)}\to \eta_c\pi^+\pi^-$
decays can proceed mainly via quasi-two-body channels which contain scalar or vector resonant states as argued
in Refs.~\cite{Chen:2002th,Wang-2014a}.
We firstly derive the PQCD factorization formulas for the $B^0_{(s)}\to \eta_c\pi^+\pi^-$
decays with the inputs of the $S$-wave two-pion DAs.
We made an hypothesis that the leading-order hard kernel for three-body $B$ meson decays contain only
one hard gluon exchange as depicted in Fig.~\ref{fig:fig1}, where the $B^0$ or $B^0_s$ meson
transits into a pair of the $\pi^+$ and $\pi^-$ mesons through an intermediate resonance.
The Figs.~\ref{fig:fig1}(a) and \ref{fig:fig1}(b) represent the factorizable contributions, while
the Figs.~\ref{fig:fig1}(c) and \ref{fig:fig1}(d) denote the spectator contributions.

%%%%%%%%%%%%%%%%%%%%%%%%%%%%%%%%%%%%%%%%%%%%%%%%%%
%%%%%%%%%%%%%%%%%%%%%%%%%%%%%%%%%%%%%%%%%%%%%%%%%%

In the light-cone coordinates, we assume that the light final state ``two pions" and $\eta_c$ is moving along
the direction of $n_+=(1,0,0_\rmt)$ and $n_-=(0,1,0_\rmt)$, respectively. The $B^0_{(s)}$ meson
momentum $p_{B}$, the total momentum of the two pions, $p=p_1+p_2$, and the $\eta_c$ momentum $p_3$ are chosen as
\beq\label{mom-pBpp3}
p_{B}=\frac{m_{B}}{\sqrt2}(1,1,0_\rmt),~\quad p=\frac{m_{B}}{\sqrt2}(1-r^2,\eta,0_\rmt),~\quad
p_3=\frac{m_{B}}{\sqrt2}(r^2,\bar{\eta},0_\rmt),
\eeq
where $m_{B}$ denotes the $B^0_{(s)}$ meson mass, the variable $\eta$ is defined as $\eta=\omega^2/[(1-r^2)m^2_{B}]$
with the mass ratio  $r=m_{\eta_c}/m_{B}$,  the variable $\etab=1-\eta$ and the invariant mass
squared $\omega^2=p^2=m^2(\pi^+\pi^-)$ of the pion pair.
As shown in Fig.~\ref{fig:fig1}(a), the momentum $k_B$ of
the spectator quark in the $B$ meson, the momentum $k=z p^+$ and $k_3=x_3 p_3$
are of the form of
\beq
\label{mom-B-k}
k_{B}&=& \left(0,\frac{m_{B}}{\sqrt2} x_B,k_{BT}\right),\quad
k=zp^+=\left(\frac{m_{B}}{\sqrt2} z(1-r^2),0,k_\rmt\right),\non
k_3&=&x_3p_3=\left(\frac{m_{B}}{\sqrt2} r^2x_3,\frac{m_{B}}{\sqrt2} \etab x_3,k_{3\rmt}\right),
\eeq
where the momentum fraction $x_{B}$, $z$ and $x_3$ run between zero and unity.

The  wave function of $B/B_s$ meson can be written as~\cite{Keum:2000ph,li2003,prd65}
\beq
\Phi_B= \frac{i}{\sqrt{2N_c}} (\psl_B +m_B) \gamma_5 \phi_B ({\bf k_1}).
\label{bmeson}
\eeq
Here we adopt the B-meson distribution amplitude $\phi_B(x,b)$ in the PQCD approach widely used
since  2001~\cite{Keum:2000ph,li2003,prd65}
\beq
\phi_B(x,b)&=& N_B x^2(1-x)^2\mathrm{exp} \left
 [ -\frac{M_B^2\ x^2}{2 \omega_{B}^2} -\frac{1}{2} (\omega_{B}\; b)^2\right],
 \label{phib}
\eeq
where the normalization factor $N_B$ depends on the value of $\omega_B$ and $f_B$ and defined through the normalization relation
$\int_0^1dx \; \phi_B(x,b=0)=f_B/(2\sqrt{6})$. $\omega_B$ is a free parameter and we take $\omega_B = 0.40 \pm
0.04$ GeV and $\omega_{B_s}=0.50 \pm 0.05$ GeV in the numerical calculations.

For the pseudoscalar meson $\eta_c$, its wave function can be written as
\begin{eqnarray}
\Psi_{\eta_c}=\frac{1}{\sqrt{2N_c}}{\gamma_5}\left[\psl_3\psi_v+m_{\eta_c}\psi_s\right],
\end{eqnarray}
here the twist-2 distribution amplitude $\psi_v$ and the twist-3 distribution amplitude
$\psi_s$ take the form of \cite{Bondar:2004sv}
\begin{eqnarray}
\psi_v(x)&=&9.58\frac{f_{\eta_c}}{2\sqrt{2N_c}}x(1-x)\left[\frac{x(1-x)}{1-2.8x(1-x)}\right]^{0.7},\non
\psi_s(x)&=&1.97\frac{f_{\eta_c}}{2\sqrt{2N_c}}\left[\frac{x(1-x)}{1-2.8x(1-x)}\right]^{0.7},
\end{eqnarray}
where $f_{\eta_c}$ is the decay constant of $\eta_c$ meson.

The $S$-wave two-pion distribution amplitude $\Phi_{\pi\pi}^{\rm S}$ have been defined in
Ref.~\cite{Meissner:2013hya}
\beq
\Phi_{\pi\pi}^{\rm S}=\frac{1}{\sqrt{2N_c}}\left[\psl\Phi_{v\nu=-}^{I=0}(z,\zeta,\omega^2)
+\omega\Phi_{s}^{I=0}(z,\zeta,\omega^2)+\omega(\nsl_+\nsl_--1)\Phi_{t\nu=+}^{I=0}(z,\zeta,\omega^2) \right],
\label{eq:phifunc}
\eeq
where $\zeta=p_1^+/p^+$ is the momentum fraction of the $\pi^+$ in the pion pair,
the asymptotic forms of the individual DAs in Eq.~(\ref{eq:phifunc}) have been parameterized
as~\cite{MP,MT,MN}
\begin{eqnarray}
\Phi_{v\nu=-}^{I=0}&=&\phi_0=\frac{9F_{s}(\omega^2)}{\sqrt{2N_c}}a_2^{I=0}z(1-z)(1-2z),\non
\Phi_{s}^{I=0}&=&\phi_s=\frac{F_{s}(\omega^2)}{2\sqrt{2N_c}},\non
\Phi_{t\nu=+}^{I=0}&=&\phi_t=\frac{F_{s}(\omega^2)}{2\sqrt{2N_c}}(1-2z),
\end{eqnarray}
with the time-like scalar form factor $F_{s}(\omega^2)$ and the Gegenbauer coefficient $a_2^{I=0}$.
For simplicity, we here denote the distribution
amplitudes $\Phi_{v\nu=-}^{I=0}(z,\zeta,\omega^2)$, $\Phi_{s}^{I=0}(z,\zeta,\omega^2)$
and $\Phi_{t\nu=+}^{I=0}(z,\zeta,\omega^2)$ by $\phi_0$, $\phi_s$ and $\phi_t$, respectively.

Following the LHCb collaboration ~\cite{lhcb02,Aaij:2014emv,lhcb01}
\footnote{In  their analysis \cite{lhcb02,Aaij:2014emv,lhcb01},
the LHCb collaboration used the Flatt\'{e} model \cite{F-model} for the description of $f_0(980)$, the Breit-Wigner model
for $f_0(1500)$ and $f_0(1790)$. },
we also introduce the $S$-wave resonances into the parametrization of the function $F_{s}(\omega^2)$,
so that both resonant and non-resonant contributions are included into the $S$-wave
two-pion wave function $\Phi_{\pi\pi}^{\rm S}$ .
For the $s\bar s$ component in the $B_s\to \eta_c\pi^+\pi^-$
decay, we take into account the contributions from the intermediate resonant $f_0(980),
f_0(1500)$  and $f_0(1790)$ as in Ref.~\cite{Aaij:2014emv}.
We use the Flatt\'{e} model \cite{F-model} for $f_0(980)$ as given in Eq.~[18] of  Ref.~\cite{Aaij:2014emv}, and the
Breit-Wigner model for $f_0(1500)$ and $f_0(1790)$.
The $s\bar s$ component of the time-like scalar form factor $F_{s}(\omega^2)$, consequently,  can be written as
the form of
\beq
F_s^{s\bar s}(\omega^2)&=&
\frac{c_1m_{f_0(980)}^2e^{i\theta_1}}{m_{f_0(980)}^2-\omega^2-im_{f_0(980)}
  (g_{\pi\pi}\rho_{\pi\pi}+g_{KK}\rho_{KK})} \nonumber\\
  &+&\frac{c_2m_{f_0(1500)}^2e^{i\theta_2}}{m_{f_0(1500)}^2-\omega^2-im_{f_0(1500)}\Gamma_{f_0(1500)}(\omega^2)}\non
&+&\frac{c_3m_{f_0(1790)}^2e^{i\theta_3}}{m_{f_0(1790)}^2-\omega^2-im_{f_0(1790)}\Gamma_{f_0(1790)}(\omega^2)},
\label{eq:fsss}
\eeq
here the three terms describe the contributions from $f_0(980)$, $f_0(1500)$, and $f_0(1790)$, respectively.
All relevant parameters in above equation are the same as those being defined previously
in Refs.~\cite{Aaij:2014emv,pdg2014,Wang-2015a}, such as
\beq
m_{f_0(980)}&=&0.97 {\rm GeV}, \quad g_{\pi\pi}=0.167, \quad g_{KK}=3.47 g_{\pi\pi}, \non
m_{f_0(1500)} &=& 1.50 {\rm GeV}, \quad m_{f_0(1790)} = 1.81 {\rm GeV}.
\eeq
We assume that the energy-dependent width $\Gamma_S(\omega^2)$ for a $S$-wave resonance decaying into two
pions is parameterized in the same way as in Ref.~\cite{prl86-216}
\beq
\Gamma_S(\omega^2)=\Gamma_S \; \frac{m_S}{\omega}\;
\left(\frac{\omega^2-4m_{\pi}^2}{m_S^2-4m_{\pi}^2} \right)^{\frac{1}{2}} F_R^2,
\eeq
with the pion mass $m_{\pi}=0.13$ {\rm GeV}, the constant width $\Gamma_S$ with $\Gamma_S =0.12, 0.32$ {\rm GeV} for
$f_0(1500)$ and $f_0(1790)$ respectively, and the Blatt-Weisskopf barrier factor $F_R=1$ in this case~\cite{lhcb02}.

For the $d\bar d$ component $F_s^{d\bar d}(\omega^2)$, only the resonance
$f_0(500)$ or the so-called $\sigma$ meson in literature is relevant.
Because the resonance $f_0(500)$
is complicated and has a wide width, we here parameterize the $f_0(500)$ contribution to the scalar
form factor for the $d\bar d$ component in two different ways: the Breit-Wigner and the D.~V.~Bugg model \cite{DVBugg:2007},
respectively.
Following Refs.~\cite{lhcb01,pdg2014}, we firstly adopt the Breit-Wigner model with the pole mass
$m_{f_0(500)}=0.50$ {\rm GeV} and the width $\Gamma_{f_0(500)}=0.40$ {\rm GeV},
\beq
F_s^{d\bar d}(\omega^2)=\frac{c m_{f_0(500)}^2}{m_{f_0(500)}^2-\omega^2
-im_{f_0(500)}\Gamma_{f_0(500)}(\omega^2)}. \label{eq:fsdd}
\eeq
The parameters $c$, $c_i$ and $\theta_i$ with $i=(1,2,3)$ appeared in Eqs.~(\ref{eq:fsss}) and (\ref{eq:fsdd})
have been extracted from the LHCb data~\cite{Aaij:2014emv},
\beq
c_1&=&0.900,\quad c_2=0.106, \quad c_3=0.066, \quad c=3.500,\non
\theta_1&=&-\frac\pi2, \quad~~ \theta_2=\frac\pi4, \quad\quad~ \theta_3=0.
\label{para}
\eeq
Secondly, we parameterize the form factor of $f_0(500)$ with the D.~V.~Bugg~ resonant lineshape \cite{DVBugg:2007}
in the same way as in Ref.~\cite{Aaij:2015}
\beq
T_{11}(s) &=& M \; \Gamma _1(s)   \left [ M^2 - s - g^2_1
\frac {s-s_A}{M^2-s_A} \left [ j_1(s) - j_1(M^2) \right ] - iM \sum_{i=1}^4 \Gamma_i(s) \right ]^{-1}, \label{eq:bugg}
\eeq
where $s=\omega^2 =m^2(\pi^+\pi^-)$, $j_1(s) = \frac {1}{\pi}\left[2 + \rho _1  \ln \left( \frac {1 - \rho _1}{1
+ \rho _1}\right) \right]$,
the functions $g^2_1(s)$,$\Gamma_i(s)$ and other relevant functions in Eq.~(\ref{eq:bugg} ) are the following
\beq
g^2_1(s) &=& M(b_1 + b_2s)\exp [-(s - M^2)/A],\non
M\; \Gamma_1(s) &=& g^2_1(s)\frac {s-s_A}{M^2-s_A}\rho_1(s), \non
M\; \Gamma_2(s) &=& 0.6g^2_1(s)(s/M^2)\exp (-\alpha |s-4m^2_K|)\rho_2(s),\non
M\;\Gamma_3(s) &=& 0.2g^2_1(s)(s/M^2)\exp (-\alpha |s-4m^2_\eta|)\rho_3(s), \non
M\;\Gamma_4(s) &=& M\; g_{4\pi}\; \rho_{4\pi}(s)/\rho_{4\pi }(M^2), \ \ with  \ \
\rho _{4\pi}(s) = 1.0/[1 + \exp (7.082 - 2.845s)], \label{eq:buggls}
\eeq
For the parameters in Eqs.~(\ref{eq:bugg},\ref{eq:buggls}), we use their values as given in the
fourth column of Table I in Ref.~\cite{DVBugg:2007}:
\beq
M &=& 0.953 {\rm GeV}, \quad s_A = 0.41 \ m_{\pi}^2, \quad b_1 = 1.302 {\rm GeV}^2,\non
b_2&=& 0.340, \quad A = 2.426 {\rm GeV}^2, \quad g_{4\pi} = 0.011 {\rm GeV}.
\label{eq:input1}
\eeq
And the parameters $\rho_{1,2,3}$ in Eq.~(\ref{eq:buggls})
are the phase-space factors of the decay channels $\pi\pi$, $KK$ and $\eta\eta$ respectively, and have been defined as
\cite{DVBugg:2007}
\beq
\rho_i(s) = \sqrt {1 - 4\frac{m^2_i }{s} },
\eeq
with $m_1=m_\pi, m_2=m_K$ and $m_3=m_\eta$.

The differential decay rate for the $B^0_{(s)}\to \eta_c\pi^+\pi^-$ decay can be written as~\cite{pdg2014}
\beq
\frac{d{\cal B}}{d\omega}=\tau_{B}\frac{\omega|\vec{p}_1|
|\vec{p}_3 | }{4(2\pi)^3m^3_{B}}|{\cal A}|^2,
\label{expr-br}
\eeq
where $\omega=m(\pi^+\pi^-)$, $|\vec{p}_1|$ and $|\vec{p}_3|$ denote the magnitudes of
the $\pi^+$ and $\eta_c$ momenta in the center-of-mass frame of the pion pair,
\beq
|\vec{p}_1|&=&\frac12\sqrt{\omega^2-4m^2_{\pi^{\pm}}}, \non
|\vec{p}_3|&=&\frac{1}{2\omega}
\sqrt{\left[m^2_{B}-(\omega+m_{\eta_c})^2 \right]\left[m^2_{B}-(\omega-m_{\eta_c})^2 \right]}.
\eeq

The decay amplitude for the decay $B^0_{(s)}\to \eta_c\pi^+\pi^-$ is of the form
\beq
\mathcal{A}({B^0_{(s)}\to \eta_c\pi^+\pi^-})&=&V^*_{cb}V_{cd(cs)}\left(F^{LL}+M^{LL}\right)\non
                                            &-&V^*_{tb}V_{td(ts)}\left(F^{\prime LL}+F^{LR} +M^{\prime LL} +M^{SP} \right),
\eeq
where the functions $F^{LL}, F^{\prime LL}$ and $F^{LR}$ ( $M^{LL}, M^{\prime LL}$ and $M^{LR}$ )
denote the amplitudes for the $B/B_s$ meson transition into two pions as illustrated by
Fig.~1(a) and 1(b)  ( Fig.~1(c) and 1(d)):
\beq
F^{LL}&=&8\pi C_F m^4_B f_{\eta_c}\int_0^1 dx_B dz \int_0^\infty b_B\; db_B\; b\; db\; \phi_B(x_B,b_B)\non
&\times&\bigg\{\bigg[\sqrt{\eta(1-r^2)}\big[[(1-2z)\etab +r^2(1+2z \etab ) ]\phi_s
-(r^2(1-2z\etab )+(2z-1)\etab )\phi_t\big]\non
&-&\big[ -(1+z)\etab  + r^2(1+2z\etab -2\eta) -r^4z\etab \big]
\phi_0\bigg]  a_1(t_a)\; E_e(t_a)\; h_a(x_B,z,b_B,b)\non
&+&\bigg[2\sqrt{\eta(1-r^2)}\big[r^2(\eta-x_B)+(1-r^2) \etab \big]\phi_s
-(1-r^2)\big[ \eta\etab + r^2(\eta-x_B)\big] \phi_0\bigg] \non
& \times& a_1(t_b)\; E_e(t_b)\; h_b(x_B,z,b_B,b)
\bigg\}, \label{eq:fll01}\\
F^{\prime LL}&=&F^{LL}|_{a_1\to a_2},\quad  F^{LR}=-F^{LL}|_{a_1\to a_3}, \label{exp-F-LL}
\eeq
\beq
M^{LL}&=&32\pi C_F m^4_B/\sqrt{6} \int_0^1 dx_B dz dx_3 \int_0^\infty b_B db_B\; b_3 db_3\; \phi_B(x_B,b_B)\non
&\times&\bigg\{\bigg[\sqrt{\eta(1-r^2)}\Big[ \big[r^2x_B+(1-\eta)(1-r^2)z \big]\phi_t\psi_v\non
&-& \big[ 2r^2 (1-x_3) \etab +(1-r^2)z\etab -r^2x_B \big]\psi_v \phi_s -4rr_c\psi_s \phi_s \Big]\non
&+& (\etab+r^2)\big[(1-r^2)(1-x_3-x_B)+ x_3(1-2r^2)\eta -(1-r^2)(1-z)\eta +r^2\eta \big]\psi_v \phi_0 \non
&-&rr_c(1-r^2+\eta)\psi_s \phi_0\bigg]  C_2(t_c)\; E_n(t_c)\; h_c(x_B,z,x_3,b_B,b_3)\non
&+&\bigg[\sqrt{\eta(1-r^2)}\Big[ \big[2r^2x_3 \etab -r^2x_B+ z(1-r^2)\etab \big] \psi_v\phi_s
+\big[ z(1-r^2)\etab +r^2x_B \big]\psi_v\phi_t\Big]\non
&-&[ \etab +r^2(\eta-\etab) ] \big[x_3 \etab-x_B+r^2x_3 +z(1-r^2)\big]\psi_v\phi_0\bigg]\non
&\times&C_2(t_d)\; E_n(t_d)\;h_d(x_B,z,x_3,b_B,b_3)\bigg\},\\
M^{\prime LL}&=&M^{LL}|_{C_{2}\to C_4+C_{10}},
\label{exp-M-LL}
\eeq
\beq
M^{SP}&=&32\pi C_F m^4_B/\sqrt{6} \int_0^1 dx_B\; dz\; dx_3 \int_0^\infty b_B db_B\; b_3 db_3\; \phi_B(x_B,b_B)\non
&\times&\bigg\{
\bigg[\sqrt{\eta(1-r^2)}\Big[ \big[ 2r^2(1-x_3)\etab +z(1-r^2)\etab -r^2x_B \big]\psi_v\phi_s
-4rr_c\psi_s \phi_s \Big] \non
&+&\sqrt{\eta(1-r^2)}\left[z(1-r^2)\etab + r^2x_B\right]\psi_v\phi_t\non
&-&\Big[ \left (-r^2+\etab + 2r^2\eta \right ) \big[ (1-x_3)r^2+z(1-r^2)+(1-x_3)\etab -x_B \big]\psi_v \phi_0\non
&-&rr_c(1+\eta-r^2)\psi_s\phi_0 \Big]  \bigg]  \left [C_6(t_c)+C_8(t_c)\right ]\; E_n(t_c)\; h_c(x_B,z,x_3,b_B,b_3)\non
&-&\bigg[ \sqrt{\eta(1-r^2)}\Big[ \big[ 2r^2x_3 \etab+(1-r^2)z\etab -r^2x_B \big] \psi_v \phi_s
-\big[ z(1-r^2)\etab +r^2x_B \big]\psi_v \phi_t\Big]\non
&-&(r^2 +\etab )\Big[(1-r^2)( x_3\etab -x_B)+x_3r^2\eta+\eta z(1-r^2)\Big]\psi_v\phi_0\bigg] \non
&\times& \left [C_6(t_d)+C_8(t_d)\right ]\; E_n(t_d)\; h_d(x_B,z,x_3,b_B,b_3)
 \bigg\},
\label{eq:msp01}
\eeq
where $C_F=4/3$, $r_c=m_c/m_B$, and $a_i$ are the combinations of the Wilson coefficients $C_i$:
\beq
a_1&=& C_1+\frac{C_2}{3}, \quad a_2=C_3+C_9+ \frac{C_4 +C_{10}}{3},\quad
a_3= C_5+C_7+\frac{C_6+C_8}{3}.
\label{eq:ai123}
\eeq
The explicit expressions of the evolution factors $(E_e(t_a),E_e(t_b),E_n(t_c),E_n(t_d))$, the
hard functions $(h_a,h_b,h_c,h_d)$ and the hard scales $(t_a,t_b,t_c,t_d)$, appeared in
Eqs.~(\ref{eq:fll01}-\ref{eq:msp01}),  can be found in the appendix of Ref.~\cite{Wang-2015a}.

For the factorizable emission diagrams Fig.~1(a) and 1(b), the NLO vertex  corrections
can be taken into account through the inclusion of additional terms to the Wilson
coefficients ~\cite{nlo05,npb675,bbns99,buras96}. After the inclusion of the NLO vertex corrections, the Wilson coefficients
$a_1$, $a_2$ and $a_3$ as defined in Eq.~(\ref{eq:ai123}) will be modified into the following form
\beq
a_1(\mu) &=& C_1(\mu) +\frac{C_2(\mu)}{3}
\left \{ 1+ \frac{2 \alpha_s(\mu) }{3\pi}   \left [ 6\ln\frac{m_b}{\mu}-9+\frac{ \sqrt{6} }{f_{\eta_c} }
\int_0^1 dx\psi_v(x) g(x) \right ] \right \}, \label{eq:a1mu}\\
a_2(\mu) &=& C_3(\mu) + C_9(\mu) +\frac{C_4(\mu)+C_{10}(\mu)}{3}\non
& \times& \left \{ 1+ \frac{2 \alpha_s(\mu) }{3\pi}  \left [ 6\ln\frac{m_b}{\mu}-9+\frac{ \sqrt{6} }{f_{\eta_c} }
\int_0^1 dx\psi_v(x) g(x) \right ] \right \}, \label{eq:a2mu}\\
a_3(\mu) &=& C_5(\mu) + C_7(\mu) +\frac{C_6(\mu)+C_{8}(\mu)}{3}\non
&\times& \left \{ 1 - \frac{2 \alpha_s(\mu) }{3\pi}   \left [ 6\ln\frac{m_b}{\mu}-3+\frac{ \sqrt{6} }{f_{\eta_c} }
\int_0^1 dx\psi_v(x) g(1-x) \right ] \right \}. \label{eq:a3mu}
\eeq
Since the emitted meson $\eta_c$ is heavy, the terms proportional to the factor $z=m^2_{\eta_c}/m^2_B$ can not be
neglected. One therefore should use the hard-scattering functions $g(x)$ as given in Ref.~\cite{zsc2004}
instead of the one in Ref.~\cite{nlo05},
\beq
g(x)&=& \frac{3(1-2x)}{1-x}\ln[x]+3 \left[  \ln(1-z)-i \pi \right]
-\frac{2z (1-x)}{1-z x}-\frac{2 x z(\ln[1 - z]-i \pi )}{1 - (1 - x) z}   \non
&-& \frac{ x^2 z^2 (\ln[1 - z]-i \pi )}{(1 - (1 - x) z)^2}
 +x z^2  \ln[x z]\left [ \frac{x}{(1-(1 - x) z)^2}-\frac{1-x}{(1-x z)^2}\right ]\non
&+& 2 x z \ln[x z] \left [ \frac{1}{1-(1 - x) z}  -\frac{1}{1-x z}\right],
\eeq
where $z=m^2_{\eta_c}/m^2_B$. For $B_s \to \eta_c \pi^+\pi^-$ decay, the mass $m_B$ in above equations should be
replaced by the mass $m_{B_s}$.

%%%%%%%%%%%%%%%%%%%%%%%%%%%%%%%%%%%%%%%%%%%%%%%%%%%%

\section{Numerical results}\label{sec:3}

In numerical calculations, besides the quantities specified before, the following input parameters (the masses, decay constants and QCD scale are in units of {\rm GeV} ) will be used ~\cite{Wang-2014a,pdg2014}
\beq
\Lambda^{(f=4)}_{ \overline{MS} }&=&0.326, \quad m_{B^0}=5.280, \quad m_{B_s}=5.367,
\quad m_{\eta_c}=2.9836, \non
m_{\pi^\pm}&=&0.140, \quad m_{\pi^0}=0.135, \quad
m_{b}=4.8, \quad m_c=1.275, \quad m_s=0.095, \non
f_B&=& 0.19\pm 0.02, \quad f_{B_s}=0.236, \quad \tau_{B^0}=1.519\; ps, \quad
\tau_{B_{s}}=1.512\; ps. \label{eq:inputs}
\eeq
The values of the Wolfenstein parameters are the same as given in Ref.~\cite{pdg2014}:
$A=0.814^{+0.023}_{-0.024}, \lambda=0.22537\pm 0.00061$, $\bar{\rho} = 0.117\pm0.021,
\bar{\eta}= 0.353\pm 0.013$. For the Gegenbauer coefficient we use $a_2^{I=0}=0.2$.

\begin{figure}[tb]
\begin{center}
\vspace{-0.5cm}
\centerline{\epsfxsize=8cm \epsffile{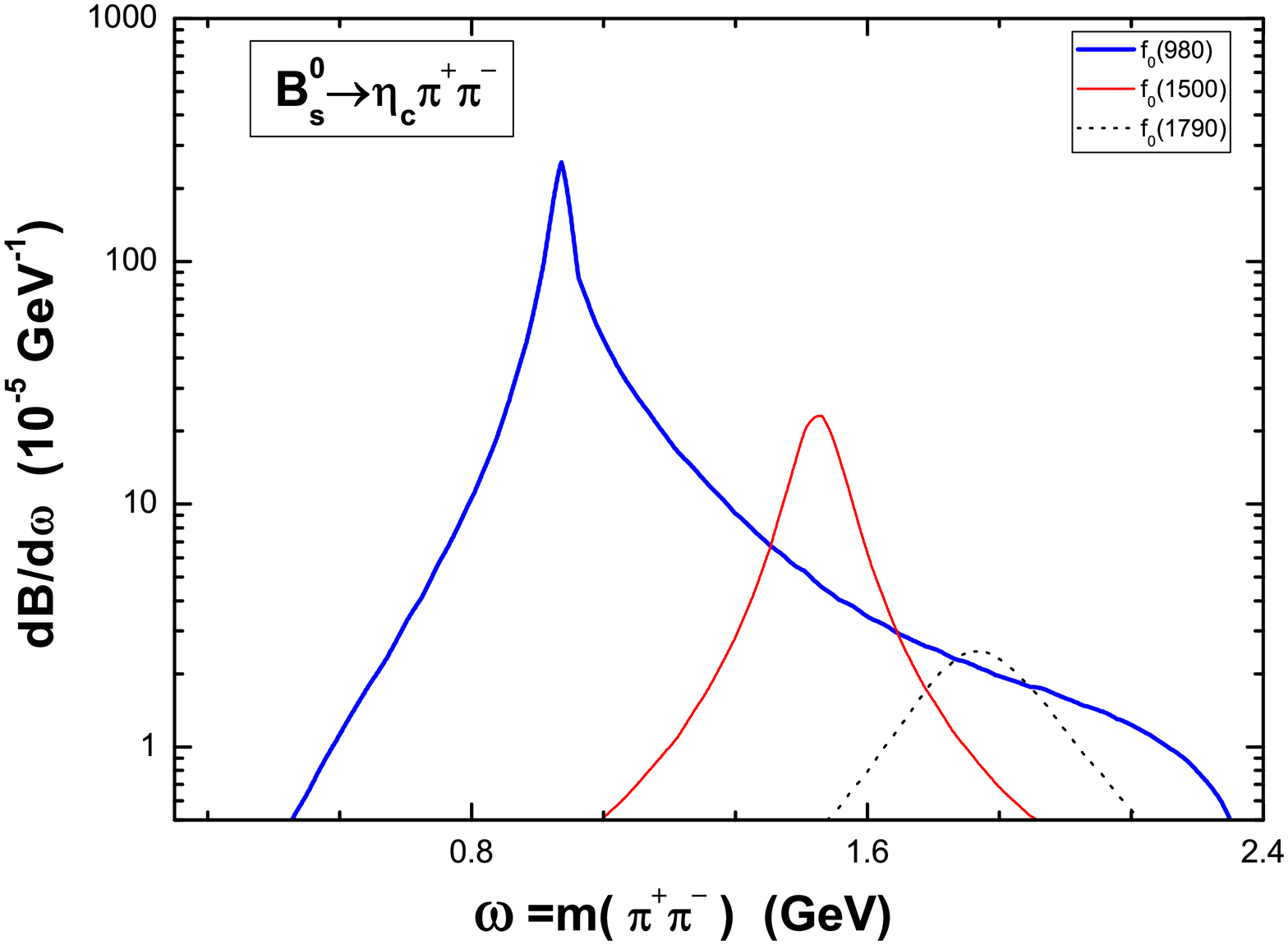}
\hspace{0.4cm}\epsfxsize=8cm \epsffile{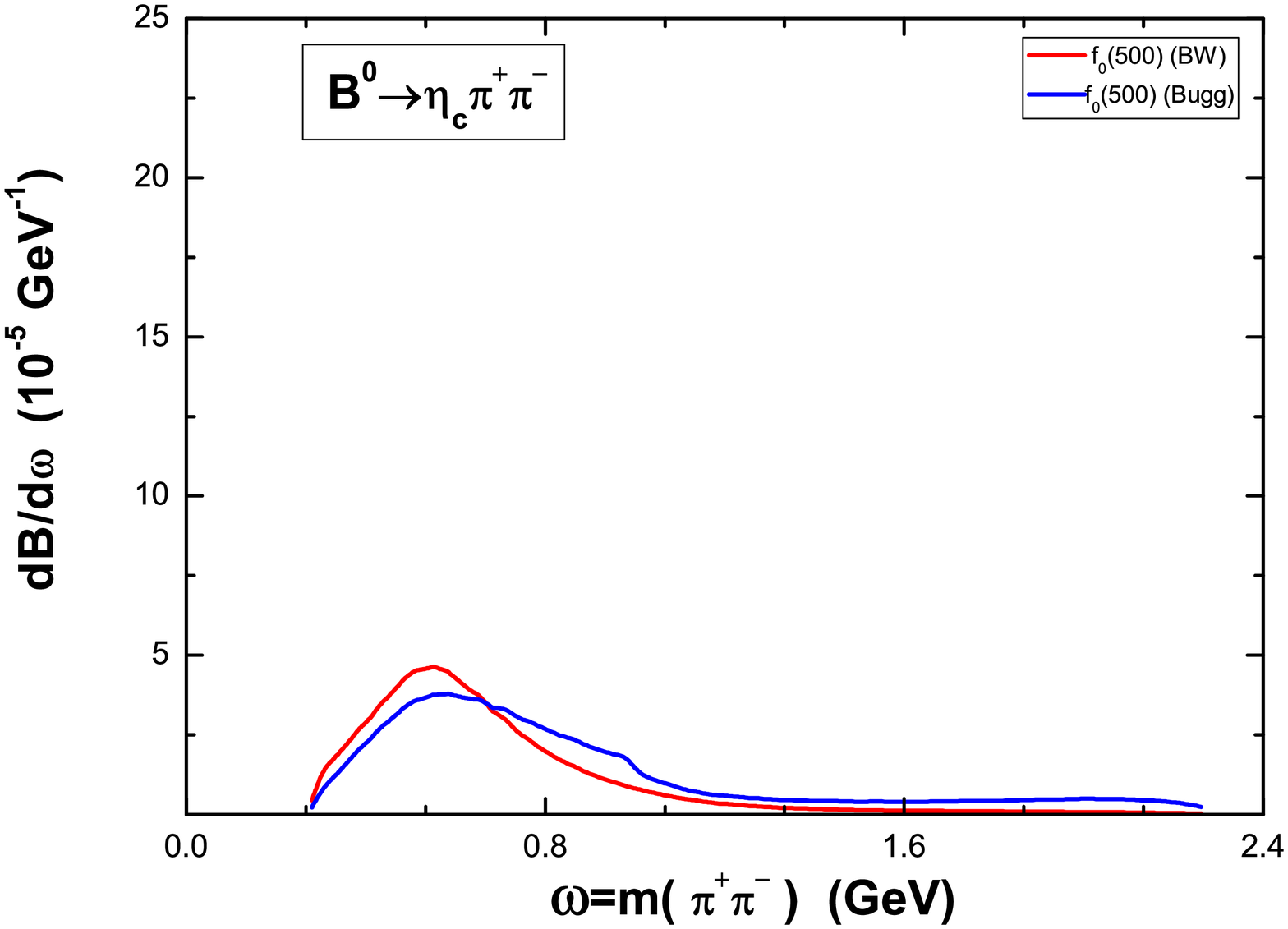}  }
(a)\hspace{8cm}(b)
\caption{ The $m(\pi^+\pi^-)$-dependence  of the differential decay rates $d{\cal B}/d\omega$
for (a) the contribution from the resonance $f_0(980)$, $f_0(1500)$ and $f_0(1790)$
to  $B^0_s\to \eta_c\pi^+\pi^-$ decay, and (b) the contribution from $f_0(500)$
to  $B^0\to \eta_c\pi^+\pi^-$ decay in the BW model ( red curve) or the Bugg model ( blue curve) . }\label{fig:fig2}
\end{center}
\end{figure}

In Fig.~\ref{fig:fig2}(a), we show the contributions to the differential decay rate
$d{\cal B}(B_s\to \eta_c \pi^+\pi^-)/d\omega$ from each
resonance  $f_0(980)$ ( the blue solid curve), $f_0(1500)$(the red solid curve) and
$f_0(1790)$ (the dots curve), as a function of the pion-pair invariant mass $\omega=m(\pi^+\pi^-)$.
For the considered $B_s$ decay, the allowed region of $\omega$ is $4m_\pi^2 \leq \omega^2 \leq (M_{B_s}-m_{\eta_c})^2$.
In Fig.~\ref{fig:fig2}(b), furthermore, we show the contribution to the differential decay rate
$d{\cal B}(B^0\to \eta_c \pi^+\pi^-)/d\omega$ from the resonance $f_0(500)$
as a function of $m(\pi^+\pi^-)$ too, where the red and blue line shows the prediction obtained by using the
Breit-Wigner model and the D.~V.~Bugg model respectively.
For $B\to \eta_c \pi^+\pi^-$ decay, the dynamical limit on the value of $\omega$ is
$4m_\pi^2 \leq \omega^2 \leq ( M_B-m_{\eta_c} )^2$.

%%-----------------------------------------------
%%-----------------------------------------------

For the decay $B \to \eta_c f_0(500) \to \eta_c \pi^+\pi^-$, the PQCD prediction
for its branching ratio with Breit-Wigner form is
\beq
{\cal B}(B^0\to \eta_c f_0(500)[\pi^+\pi^-])&=& \left[ 1.53
^{+0.44}_{-0.35}(\omega_{B})^{+0.62}_{-0.00}(a^{I=0}_2)^{+0.00}_{-0.01}(m_c)\right ]\times 10^{-6}.
\label{eq:br011}
\eeq
When we use the method of D.~V.~Bugg, the PQCD prediction for its branching ratio is of the form
\beq
{\cal B}(B^0\to \eta_c f_0(500)[\pi^+\pi^-])&=& \left[ 2.31
^{+0.63}_{-0.48}(\omega_{B})^{+0.73}_{-0.00}(a^{I=0}_2)^{+0.00}_{-0.01}(m_c)\right ]\times10^{-6},
\label{eq:br011}
\eeq
where the three major errors are induced by the uncertainties of $\omega_{B}=(0.40\pm0.04)$ {\rm GeV},
$a_2^{I=0}=0.2\pm 0.2$ and $m_c=(1.275\pm0.025)$ {\rm GeV}, respectively.

For the decay mode $B_s \to \eta_c f_0(X) \to \eta_c (\pi^+\pi^-)_S$,
when the contribution from each resonance $f_0(980)$, $f_0(1500)$ and $f_0(1790)$ are included
respectively, the PQCD predictions for the branching ratios  for each case are the following,
\beq
{\cal B}(B_s\to \eta_c f_0(980)[\pi^+\pi^-])&=&\left[3.37
^{+0.98}_{-0.77}(\omega_{B_s})^{+0.38}_{-0.00}(a^{I=0}_2)^{+0.03}_{-0.00}(m_c)\right]\times10^{-5},
\label{eq:br021} \\
{\cal B}(B_s\to \eta_c f_0(1500)[\pi^+\pi^-])&=&\left[6.76
^{+1.50}_{-1.21}(\omega_{B_s})^{+0.60}_{-0.00}(a^{I=0}_2)^{+0.09}_{-0.00}(m_c)\right]\times10^{-6},
\label{eq:br022}\\
{\cal B}(B_s\to \eta_c f_0(1790)[\pi^+\pi^-])&=&\left[1.97
^{+0.59}_{-0.44}(\omega_{B_s})^{+0.21}_{-0.00}(a^{I=0}_2)^{+0.01}_{-0.01}(m_c)\right]\times10^{-6},
\label{eq:br023}
\eeq
where the three major errors are induced by the uncertainties of $\omega_{B_s}=(0.50\pm 0.05)$ {\rm GeV},
$a_2^{I=0}=0.2\pm 0.2$ and $m_c=(1.275\pm0.025)$ {\rm GeV}, respectively.
The errors induced by the variations of the Wolfenstein parameters and other inputs are very small and have been
neglected. If we take into account the interference between different scalars $f_0(X)$, we found the
total branching ratio:
\beq
{\cal B}(B_s\to \eta_c f_0(X)[\pi^+\pi^-] )=\left[5.02^{+1.37}_{-1.08}
(\omega_{B_s})^{+0.58}_{-0.00}(a^{I=0}_2)^{+0.03}_{-0.02}(m_c)\right]\times10^{-5}.
\label{eq:br2tot}
\eeq
The interference between $f_0(980)$ and $f_0(1500)$, $f_0(980)$ and $f_0(1790)$, as well as
$f_0(1500)$ and $f_0(1790)$, will provide a contribution of $3.29\times10^{-6}, 5.59\times10^{-6}$ and
$-1.16\times10^{-6}$ to the total decay rate, respectively.

From the curves in Fig.~\ref{fig:fig2} and the PQCD predictions for the decay rates as given in
Eqs.~(\ref{eq:br011}-\ref{eq:br2tot}), one can see the following points:
\begin{enumerate}
\item[(i)]
For  $B_s \to \eta_c f_0(X) \to \eta_c (f_0(X)\to \pi^+\pi^-) $ decay, as illustrated clearly by Fig.~2(a),
the contribution from the resonance $f_0(980)$ is dominant ($\sim 70\%$),
while the contribution from $f_0(1790)$ is very small ($\sim 4\%$ only).
The interference between $f_0(980)$ and $f_0(1500)$, as well as $f_0(980)$ and $f_0(1790)$,
are constructive and can provide $\sim 20\%$ enhancement to the total decay rate.
The interference between $f_0(1500)$ and $f_0(1790)$, however, is destructive, but very
small (less than $-2\%$) in size.

\item[(ii)]
For $B \to \eta_c f_0(500) \to \eta_c \pi^+\pi^-$ decay, the PQCD predictions for its branching ratios
are around $2\times 10^{-6}$ in magnitude when we use the Breit-Wigner or the D.~V.~Bugg model to parameterize
the wide $f_0(500)$ meson. The model-dependence of the differential
decay rate $dB/d\omega (B \to \eta_c f_0(500)(\pi^+\pi^-))$,
as illustrated in Fig.~\ref{fig:fig2}(b), are indeed not significant.
Although the central value of PQCD predictions based on the D.~V.~Bugg model are moderately larger
than the one from the Breit-Wigner model, but they are still consistent within errors.

\item[(iii)]
The  $B/B_s \to \eta_c f_0(X) \to \eta_c (\pi^+\pi^-)$ decays are similar in nature
with the decays  $B/B_s \to \jpsi f_0(X) \to \jpsi (\pi^+\pi^-)$ studied previously in Ref.~\cite{Wang-2015a}.
We find numerically ${\cal B}(B \to \eta_c \pi^+\pi^-):{\cal B}(B \to \jpsi \pi^+\pi^-)\approx 0.3:1$.

\end{enumerate}

\section{Summary}\label{sec:4}

In this paper, we studied the contributions from the $S$-wave resonant
states $f_0(X)$ to the $B^0_{(s)}\to \eta_c\pi^+\pi^-$ decays by employing
the PQCD factorization approach. We calculated the
differential decay rates and the branching ratios of the decay $B^0\to \eta_c f_0(500)\to \eta_c (\pi^+\pi^-)$, the decays
$B^0_{s}\to \eta_c f_0(X) \to \eta_c (\pi^+\pi^-)$ with $f_0(X)=f_0(980), f_0(1500)$ and $f_0(1790)$
respectively. By using the $S$-wave two-pion wave function $\Phi_{\pi\pi}^{\rm S}$ the resonant
and non-resonant contributions to the considered decays are taken into account.
The NLO vertex corrections  are also included through the redefinition of the relevant Wilson coefficients.

From analytical analysis and numerical calculations we found the following points:
\begin{enumerate}
\item[(i)] For the branching ratios, we found
\beq
{\cal B}^{BW}(B^0\to \eta_c f_0(500) \to \eta_c \pi^+\pi^-)&=& \left ( 1.53 ^{+0.76}_{-0.35} \right ) \times 10^{-6}, \\
{\cal B}^{Bugg}(B^0\to \eta_c f_0(500)\to \eta_c \pi^+\pi^-)&=& \left(2.31^{+0.96}_{-0.48}\right)\times10^{-6},\\
{\cal B}(B_s\to \eta_c f_0(X)\to \eta_c \pi^+\pi^-] ) &=&\left ( 5.02^{+1.49}_{-1.08} \right )\times 10^{-5},
\eeq
where the individual errors have been added in quadrature. For the decay rate
${\cal B}(B_s\to \eta_c \pi^+\pi^-)$, the contribution from the resonance $f_0(980)$ is dominant.

\item[(ii)]
For $B \to \eta_c f_0(500) \to \eta_c \pi^+\pi^-$ decay, we used the Breit-Wigner and the D.~V.~Bugg model
to parameterize the wide $f_0(500)$ meson respectively but found that the model-dependence of the PQCD predictions
are not significant.

\item[(iii)]
The considered decays with the branching ratio at the order of $10^{-6}\sim 10^{-5}$
could be measured at the ongoing LHCb experiment. The formalism of two-hadron distribution
amplitudes, consequently, could be tested by such experiments.

\end{enumerate}

%---------- appendix ---------------------------------------%

\begin{acknowledgments}
Many thanks to Hsiang-nan Li, Cai-Dian L\"u, Wei Wang and Xin Liu for valuable discussions.
This work was supported by the National Natural Science
Foundation of China under the No.~11235005 and No.~11547038.

\end{acknowledgments}

%========================= reference=========================%

\end{document}